\documentclass[12pt,english]{article}
\usepackage[T1]{fontenc}
\usepackage[latin9]{inputenc}
\usepackage{natbib}
\usepackage{csquotes}
\usepackage{setspace}
\usepackage{graphicx}
\usepackage{mathrsfs}
\usepackage{amsmath}
\usepackage{amssymb}
\usepackage{babel}
\usepackage[text={16.5cm,23cm},centering]{geometry}
\expandafter\def\expandafter\quote\expandafter{\quote\small}

\makeatletter

\@ifundefined{date}{}{\date{}}
\makeatother

\begin{document}

\title{\textbf{Infinities as natural places}}

\author{Juliano C. S. Neves%
\thanks{nevesjcs@ime.unicamp.br%
}}

\maketitle

\begin{center}
{\it{Universidade Estadual de Campinas (UNICAMP),\\
 Instituto de Matemática, Estatística e Computação Científica, Campinas, SP, Brazil}}
\end{center}

\vspace{0.5cm}

\begin{abstract}
It is shown that a notion of natural place is possible within modern physics. For Aristotle, the elements---the primary components of the world---follow to their natural places in the absence of forces. On the other hand, in general relativity, the so-called Carter-Penrose diagrams offer a notion of end for objects along the geodesics. Then, the notion of natural place in Aristotelian physics has an analog in the notion of conformal infinities in general relativity. 

\end{abstract}


\section{Introduction}

Alongside Plato's works, Aristotle's philosophy is the most influential philosophical system created in ancient Greece. The master of the Lyceum built seminal works in all areas of knowledge. In natural sciences, Aristotelian physics was \enquote{ruled out} only in modernity, or modern era, by Newtonian physics. The main difference between Newtonian and Aristotelian physics may be found in the cause of the motion of bodies. For \citet[$319b30$]{Generation}, change in general (spatial, qualitative, quantitative, and substantial)  is produced by causes.\footnote{\cite{Schummer} presents a brief introduction to the main points of Aristotle's physics.} There are four types of cause for the Greek philosopher: material, formal, efficient, and final. In Aristotle, the motion of bodies is always described by a final cause. The final cause provides a purpose for the motion. On the other hand, in Newtonian physics, the notion of final cause for the motion of bodies is rejected.\footnote{However \enquote{the religious Newton} assumed God's final causes: \enquote{We know him only by his most wise and excellent contrivances of things, and final causes} \citep[Book III, p. 506]{Newton}.} Newton's laws do not adopt final causes to describe falling bodies or planets around the sun. In Aristotelian cosmic view, the notion of final cause is completely related to reality, which may be thought in two senses: in terms of either actuality or potentiality. As we will see, this distinction indicates a possible reason for motion, especially for the natural motion.  

For Aristotle, the motion of bodies depends on the constitution of bodies. Every body is constituted by primordial elements. Following Empedocles, for Aristotle there are four sublunar elements (earth, water, air and fire). But the Stagirite also considered the fifth element: ether, which occupies the world \enquote{above} the moon. According to Aristotle, each element has its natural place, or proper place. In the absence of forces a body, for example, consisting mainly of a larger number of a specific element, moves to the natural place of the major element that constitutes that body. This is the definition of natural motion in Aristotelian physics. On the other hand, for Newton, motion is possible only by a force as cause. Even the inertial motion had a force as cause. A cause, or force, in this Newtonian sense is similar to Aristotelian efficient cause. However, as I said, the final cause is absent in Newtonian physics.   

Then, the teleological point of view is fundamental in Aristotle. As well as for Plato, there are motivations for every event in nature within Aristotelian physics. The teleological belief, according to which every change is guided by a final cause, is present both in Plato and Aristotle. In \textit{Timaeus}, for example, Plato considers the entire cosmos as something created by Demiurge. And the cosmos, or the order of the universe, created by God-Demiurge, according to Plato, has as final cause the Good. Plato wrote that God \enquote{was good, and therefore not jealous, and being free from jealousy he desired that
all things should be like himself. Wherefore he set in order the visible world, which he found in disorder} \citep[$29e$]{Plato}. Apparently,\footnote{Apparently, because there are vestiges of teleology in the principle of least action, for example. See \cite{Stoltzner}.} the teleological point of view was banned in modern science. Indeed, according to \citet[III, \S 26]{Nietzsche2}, the modern era rejects \enquote{all teleology,} and the notion of final causes are \enquote{superfluous teleological principles} \citep[\S 13]{Nietzsche}. The motion of bodies is not guided by a final cause even in general relativity. But, as we will see, the notion of \enquote{natural place} appears in general relativity by using Carter-Penrose diagrams, or conformal diagrams \citep{Carter,Penrose}. The geodesic motion leads bodies to their \enquote{natural places} in general relativity. And in Einstein's theory, a new notion of natural place is available: the conformal infinities.

There are even more points of disagreement between ancient physics and modern one. If we consider general relativity as a modern theory (also in the sense of constructed in modern era), there will be notable differences. In general relativity, it is possible to obtain solutions of Einstein's field equations without matter. For example, Minkowski spacetime is a vacuum solution in general relativity. Moreover it is flat, i.e., there is no spacetime curvature, or gravitational phenomenon, in this important solution of Einstein's field equations. An empty space is something absurd in Aristotelian physics. This is the reason for Aristotle rejecting the notion of voids from the atomists.\footnote{See \textit{Physics III}.} Besides, a world without the notions of light or heavy bodies---which lead to the notion of gravitational phenomenon for Aristotle---is unthinkable.  

As discussed in \textit{On the heavens}, for the Stagirite the cosmos is spherical and spatially finite. In the modern cosmological model, based on general relativity, specifically the Friedmann-Lemaître-Robertson-Walker metric---which is another solution of Einstein's field equations---our universe possesses spherical symmetry, and the observable universe is finite. However, our standard cosmological model forbids the anthropocentric (or geocentric) Aristotelian model because the cosmological principle assumes that the universe is the same for different observers in different planets and galaxies. Above all, the cosmological principle agrees with modern values. Isonomy is an important value in the democratic modernity. That is, modernity assumes equality among citizens and observers. Nietzsche criticized this belief pointing out that the modern speech \enquote{everywhere, equality before the law, --- in this respect, nature is no different and no better off than we are}\footnote{\citet[\S 22]{Nietzsche}.} is just an interpretation. Thus, isonomy for observers in different places in the universe is a modern belief,\footnote{Argumentation developed in \cite{Neves4,Neves3} by using Nietzschean tools.} but it is essential to the standard cosmology with its predictions and simulations. 

Another disagreement is found in the gravitational waves experiment. In \cite{Neves} it is shown that the constructed sound from gravitational waves may be considered a point contrary to Aristotle's world view. The master of the Lyceum denied the Pythagorean notion of cosmic sound generated by celestial bodies in motion. However, LIGO Scientific Collaboration \citep{LIGO} produced sound using gravitational waves frequencies from the merger of two distant black holes. But that chapter illustrates an important feature of Aristotle that approaches Aristotelian physics and modern science: the importance of observation to construct natural science. In spite of the Aristotelian rejection of the cosmic sound, the argument used by Aristotle was observational one: \enquote{There is not only the absurdity of our hearing nothing (...) but also the fact that no effect other than sensitive is produced upon us} \citep[$290b30$]{Heavens}. Contrary to Plato and the Pythagoreans, Aristotle was not a pure mathematician or an obstinate metaphysician. 

As we saw, there are several points of disagreement between Aristotle and our modern physics. However, the dialogue between our science and ancient thought is valid. Such a dialogue may cast light upon present problems and interpretations from modern science. The suspicion of a final cause in modern physics remains. Other ancient concepts deserve attention. In this article, I will suggest that the notion of natural place has a \enquote{correspondence} in general relativity, using Carter-Penrose diagrams of spacetime. Then, an analogy may be suggested. But as we will see, like every analogy, it has limitations.

\section{Natural places in Aristotelian physics}  
The notion of place plays an important role in Aristotelian world view. In the Stagirite's interpretation, the notion of place---contrary to modern science---is most importantly than the concept of space.\footnote{See, for example, \citet[chapter 1]{Jammer}.} According to Aristotle, the concept of place must satisfy the following six requirements:
\begin{quote}
(1) that place should be the first thing surrounding that of which it is the place; and (2) not anything pertaining to the object; (3) that the primary [place] should be neither less nor greater (than the object); (4) that it should be left behind by each object [when the object moves] and be separable [from it]; further, (5) that every place should have \enquote{above} and \enquote{below}; and (6) that each body should naturally move to and remain in its proper places, and this it must do either above or below \citep[$210b32$]{Physics}. 
\end{quote}  
As we can see, the place of an object is not part of the object that occupies the place. Place, \enquote{the first thing surrounding,} is the limit, or boundary, of the surrounding body. In another important passage, this is clearly indicated: place is \enquote{the limit of the surrounding body, at which it is in contact with that which is surrounded} \citep[$212a5$]{Physics}. That which is surrounded is the body that occupies the place. Then it is necessary another object, or the surrounding body, for the definition of place of a body. That is, Aristotle's cosmic view is organic, bodies are not thought of as isolated particles. Above all, it is a whole view, which considers the whole universe as something integrated.\footnote{One may consider Einstein's attempts of adopting Mach's principle in general relativity as a modern form of an integrated cosmos.} Another interesting requirement is that places have fixed locations. Due to his cosmological interpretation, Aristotle emphasizes the place immobility in the requirement  number four. Thus, the notions of absolute above and below are presented in the number five. The requirement (6) introduces the notion of natural (or proper) place of bodies. Such a notion is motive of intense debate among scholars until today. Some interpreters suggest that natural place is some sort of cause. However, \cite{Machamer}, for example, defends that natural place is not cause (formal, material, efficient or final) of the body's natural movement. This interpretation is indicated in the fourth part of  \textit{Physics}. Place, according to Aristotle, \enquote{is not an explanation as material of things that are, for nothing is composed of it; nor as a form and definition of things; nor as an end; nor does it change things that are} \citep[$209a20$]{Physics}.

At the natural place, bodies (made up of elements) do not act on themselves. There is no interaction among bodies at the natural place. In modern science language, we may consider that the potential energy vanishes when bodies are in their natural places. That is, 
\begin{quote}
each thing moves to its own place, and this is reasonable; for what is next to something and in contact (not forced) with it, is of the same kind-if they are fused, they are not capable of being acted on (...) Again, everything remains naturally in its proper place, and this is not unreasonable; for so does the part, and what is in place is like a detached part in relation to the whole (...) \citep[$212b29$]{Physics}.
\end{quote}
Bodies at their natural places behave as if they were a unity. For all bodies, there is no movement at the natural place. Before Aristotle, Plato said in \textit{Timeaus} \enquote{that motion never exists in what is uniform} \citep[$57e$]{Plato}. Being uniformly occupied by bodies, the natural place is the \textit{kingdom of rest}. However, the bodies do not stay eternally at rest in their natural places. As we saw, Aristotelian view on the cosmos considers change as becoming (or substantial change) as well. Elements may potentially change their nature. They \enquote{are transformed, none of them is able to persist in any place allotted to it by the Order} \citep[$337a14$]{Generation}. Thus fire, for example, may become water. 

But if the natural place is not the cause of natural motion, what is the reason for such a movement? The cause for the natural motion is found in the body. \textit{The elements nature is cause of the natural motion}. Using Aristotle's notions of actuality and potentiality, according to \cite{Machamer}, and \cite{Matthen}, an element is actualized at the natural place. An element manifests completely its properties only at the natural place:
\begin{quote}
the cause of an element's natural motion is said to be its nature, that is, its \textit{arche} or principle which determines the necessity of its passing from its potential state to a state of actuality and the cause which brought it into this state of potentiality (caused it to be out of its natural place) \citep[p. 382]{Machamer}.
\end{quote} 
This elements feature will be relevant to make the analogy between Aristotelian natural place and the notion of conformal infinity in general relativity. As we will see, given the spacetime geometry (or the spacetime metric), the motion of bodies and particles is characterized by mass (or \enquote{rest mass}) in the theory of relativity. And mass is a body property that comes from its principle, or its \textit{arche}. In modern physics, a field is such an \textit{arche}.

Because of the number of elements (five), the number of natural places is five as well. The element earth occupies the center of the cosmos, its natural place is \textit{accidentally}\footnote{Aristotle says that the center of the universe and the planet earth's place coincide accidentally. \enquote{It happens, however, that the centre of the earth and of the whole is the same. Thus they [the elements earth] do move to the centre of the earth, but accidentally, in virtue of the fact that the earth's centre lies at the centre of the whole} \citep[$296b15$]{Heavens}.} the center of the planet earth. The contrary element (contrary to earth), fire, occupies the limit of the sublunar world. That is, the natural place of fire is the last place in the sublunar sphere. Above that place, according to Aristotle, we have the lunar, the planets and the fixed stars spheres, the natural place of ether. The intermediary elements (between earth and fire) occupy the intermediary natural places between the earth and fire natural places. As air is lighter than water, according to the Stagirite, the air natural place is closer to the fire natural place, and the water natural place is closer to the earth natural place, \enquote{since with each step away from earth the matter manifestly becomes finer in the same proportion as water is finer than earth} \cite[$287b20$]{Heavens}.

In this sense, the gravitational phenomenon---the falling bodies---occurs because of the existence of  \enquote{heavy} elements in heavy bodies. A body with $x$ quantity of elements earth is heavier than a body with smaller number of such an element.  The notions of \enquote{above} and \enquote{below} are given by the presence of a major part of light and heavy elements in the bodies, respectively. As we saw, in Aristotelian geocentric model, \enquote{above} and \enquote{below} are not relative, are absolute. 

Thinking about the concept of geodesics, some scholars have proposed an analogy between Aristotelian physics and general relativity. Max Jammer \cite[p. 20]{Jammer}, for example, claimed that \enquote{the idea of \enquote{geodesic lines} (...) and their importance for the description of the paths of material particles or light rays suggest a certain analogy to the notion of \enquote{natural places}(...)}. For me, the notion of infinities indicated by the Carter-Penrose diagrams suggests a better analogy. Because the concept of geodesic is contained in the spacetime diagrams.

\section{Carter-Penrose diagrams and a notion of natural place}
In general relativity \citep{Einstein} one may distinguish three types of trajectories: timelike, spacelike, and null. Geometrically speaking, according to the convention adopted in this article, given a tangent vector $v^\mu$ to any trajectory, the norm of $v^\mu$ is negative for timelike trajectories, positive for spacelike trajectories, and it vanishes for null trajectories. Considering only the gravitational field, such trajectories are called geodesics, and the vector $v^\mu$ is parallel transported along the trajectory. The geodesics equation is written as
\begin{equation}
\frac{d^2 x^\mu (\lambda)}{d \lambda^2}+\Gamma_{\nu \alpha}^{\mu} \frac{d x^{\nu}(\lambda)}{d \lambda} \frac{d x^{\alpha}(\lambda)}{d \lambda}=0,
\label{Geodesic_Eq}
\end{equation}  
where $\lambda$ is the affine parameter, and $v^\mu=dx^\mu/d\lambda$. In timelike geodesics, $\lambda$ plays the role of the proper time of an observer. In a spacetime, the trajectory of a specific body, $x^\mu (\lambda)$, is solution of Eq. (\ref{Geodesic_Eq}) and depends on the spacetime metric, $g_{\mu\nu}$, which appears in the definition of the affine connection $\Gamma_{\nu \alpha}^{\mu}$:
\begin{equation}
\Gamma_{\nu \alpha}^{\mu} = \frac{1}{2}g^{\mu\beta} \left( \frac{\partial g_{\alpha \beta}}{\partial x^{\nu}}+ \frac{\partial g_{\nu\beta}}{\partial x^{\alpha}} -\frac{\partial g_{\nu\alpha}}{\partial x^{\beta}} \right).
\end{equation}
Thus, we need to know the spacetime geometry to evaluate the motion of bodies.\footnote{At least for test particles. In this case, the tiny mass of that particles will not modify the metric, or spacetime geometry. On the other hand, massive bodies in motion modify the spacetime geometry. In this case, it is necessary to use backreaction effects to evaluate the body trajectory.} 

Carter-Penrose diagrams explore the notion of geodesic motion, which is the main concept to study the spacetime structure. The spacetime structure provides the notion of causality in general relativity. Therefore, Carter-Penrose diagrams is a powerful tool to look at the notion of causality. Moreover, the Carter-Penrose diagrams cast light upon the asymptotic form of spacetime, i.e., the representation of infinity. Roughly speaking, the spacetime diagrams represent an infinite spacetime onto a finite region. The spacetime infinities are indicated by points and lines in a piece of paper. Carter-Penrose diagrams are useful in the study of gravitational waves and black hole physics. The causal structure (and its possible violations) is well-understood in such complicated diagrams. In this article, I will present the simplest Carter-Penrose diagram, from Minkowski spacetime, and the diagram of the main metric in cosmology---the Friedmann-Lemaître-Robertson-Walker metric. However, diagrams of other spacetimes are available in textbooks \citep{Hawking,Wald}. More complicated diagrams like Schwarzschild, which describes a nonrotating black hole, and Kerr, which describes a rotating black hole, are asymptotically Minkowski, or flat, at infinity. That is, for asymptotically flat spacetimes, the conformal infinities are the Minkowskian infinities. These geometries present spacetime curvature that vanishes at infinity. As we will see, the standard cosmological geometry presents similar infinities to Minkowski spacetime as well.  

\subsection{Minkowski spacetime}

Minkowski's geometry is represented by the metric\footnote{In this article, one adopts geometric unities. Then the speed of light in vacuum is set equal to 1.}
\begin{equation}
ds^2 =\eta_{\mu\nu}dx^\mu dx^\nu = - dt^2+dr^2+r^2 \left( d\theta^2+\sin^2\theta d\phi^2  \right)
\end{equation}    
in the $t$ (time), $r$ (radial coordinate), $\theta$ (polar angle), and $\phi$ (azimuthal angle) coordinates with ranges: $-\infty<t<\infty$, $0\leq r <\infty$, $0 \leq \theta \leq \pi$, and $0 \leq \phi \leq 2\pi$. As we can see, Minkowski spacetime is spatially infinite and eternal. The procedure of constructing Carter-Penrose diagrams involves coordinate transformations. Due to the metric symmetry (spherical symmetry in Minkowski spacetime and in the cosmological metric), all coordinate transformations are related only to $t$ and $r$ coordinates, then the Carter-Penrose diagrams are two-dimensional. Specifically, the coordinate transformations use functions that \enquote{bring} infinity onto a finite region. And geometrical functions such as tangent and arc-tangent are appropriate for that task. After some coordinate transformations, which are detailed in \cite{Hawking}, and \cite{Wald}, it is possible to conclude that Minkowski spacetime is conformal to the Einstein static universe.\footnote{Einstein static universe \citep{Einstein2} is a special cosmological solution of Einstein's field equations. It describes a non-expanding universe with dust, or pressureless matter. Moreover, Einstein adopted the cosmological constant to achieve such a solution of his field equations. It is well-known that Einstein static universe is unstable.} That is, both the Einstein metric and the Minkowski metric are related by
\begin{equation}
g_{\mu\nu (E)}=\Omega^2 \eta_{\mu\nu},
\end{equation}  
where the factor $\Omega^2$ is the so-called conformal factor, and $g_{\mu\nu (E)}$ is the Einstein static universe metric in the $T$, $R$, $\theta$, and $\phi$ coordinates, explicitly written as
\begin{equation}
ds^2 = g_{\mu\nu (E)}dx^\mu dx^\nu=- dT^2+dR^2+\sin^2 R \left( d\theta^2+\sin^2\theta d\phi^2  \right).
\label{Einstein}
\end{equation}
The coordinates $t$, $T$ (time), $r$, and $R$ (radial coordinate) are related by the transformations
\begin{align}
t&=\frac{1}{2}\left[\tan\left(\frac{T+R}{2} \right)+\tan\left(\frac{T-R}{2} \right)  \right] , \nonumber \\
r&=\frac{1}{2}\left[\tan\left(\frac{T+R}{2} \right) -\tan\left(\frac{T-R}{2} \right)  \right].
\label{Transformations}
\end{align}
With the transformations (\ref{Transformations}), one has explicitly $\Omega^2=(\cos T +\cos R)^2$. But the metric (\ref{Einstein}) is the Einstein static universe with limited coordinate ranges, i.e.,  
\begin{align}
-\pi & <T+R<\pi,\nonumber \\
-\pi & <T-R<\pi,\nonumber \\
0& \leq R.
\end{align} 
Thus, the entire Minkowski spacetime is described by a region of the Einstein static universe. Geometrically speaking, there exists a conformal isometry, which is a type of diffeomorphism,\footnote{A diffeomorphism is special type of map between two manifolds, which may be represented by sets. In general relativity, a spacetime is a manifold $\mathcal{M}$ endowed with a metric $g_{\mu\nu}$. It is also indicated by $(\mathcal{M}, g_{\mu\nu})$. A spacetime is a set of events. Two manifolds are diffeomorphic if there exists a differentiable and invertible map between them. See \cite[Appendix C]{Wald}.} of Minkowski spacetime into an open region $\mathcal{O}$ of  the Einstein static universe. That is, the conformal infinities in Minkowski spacetime are boundaries of $\mathcal{O}$.  

\begin{table}[t]
  \centering
  \begin{tabular}{ | c | c | c | }
    \hline
    \textbf{\footnotesize Carter-Penrose Diagram} & \textbf{\footnotesize Origin of the Trajectories} & \textbf{\footnotesize End of the Trajectories}\\ \hline
    
    \begin{minipage}{4.5cm}
     \includegraphics[width=\linewidth, height=60mm,trim=50 170 0 0]{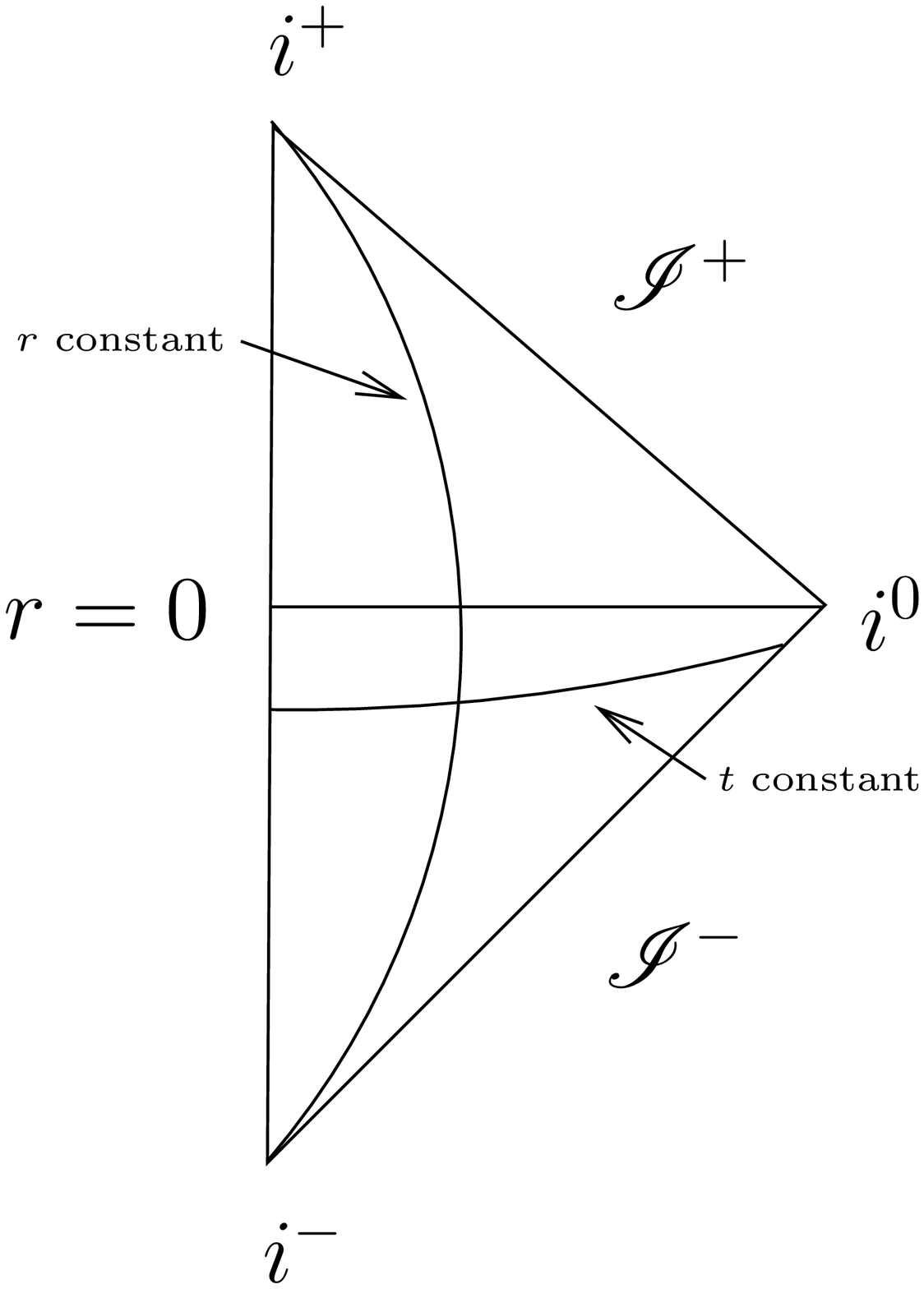}
    \end{minipage}
    &
      \begin{minipage}{5cm}
      \begin{itemize}
        \item {\scriptsize $i^{-}$  (for timelike geodesics)} 
        \item {\scriptsize $\mathscr{I}^{-}$  (for null geodesics)}
        \item {\scriptsize  $i^{0}$  (for spacelike geodesics)}
      \end{itemize}
    \end{minipage}
    & 
     \begin{minipage}{5cm}
      \begin{itemize}
        \item {\scriptsize $i^{+}$  (for timelike geodesics)}
        \item {\scriptsize $\mathscr{I}^{+}$ (for null geodesics)}
        \item {\scriptsize $i^{0}$ (for spacelike geodesics)}
      \end{itemize}
    \end{minipage}
    \\ \hline
  \end{tabular}
  \caption{Carter-Penrose diagram for Minkowski spacetime. The conformal infinities indicate the origin and end for the trajectories of bodies. The infinities $i^{-}$, past timelike infinity, $i^{+}$, future timelike infinity, and $i^{0}$, spatial infinity, are points, or vertexes, in the diagram. On the other hand, the null infinities $\mathscr{I}^{-}$, past null infinity, and $\mathscr{I}^{+}$, future null infinity, are lines inclined at $45^{\circ}$. Timelike geodesics are vertical (or almost vertical) lines, and spacelike geodesics are horizontal (or almost horizontal) lines. Null geodesics define the light cones, i.e., the spacetime causal structure. The point $r=0$ indicates the origin of the coordinate system $t$, $r$, $\theta$, and $\phi$.}
\label{Diagram_Flat}
\end{table}

In Table \ref{Diagram_Flat} one has Minkowski spacetime and its Carter-Penrose diagram. The three types of geodesics have their origins and ends represented in the diagram. For bodies along the null geodesics, the origin and end for such trajectories are $\mathscr{I}^{-}$ and $\mathscr{I}^{+}$, respectively. The line $\mathscr{I}^{-}$ is the past null infinity, and the line $\mathscr{I}^{+}$ is the future null infinity. For bodies along the spacelike geodesics, the point $i^{0}$ is the origin and end, which is called spatial infinity. Lastly, a body traveling along a timelike geodesic presents $i^{-}$, the past timelike infinity, as origin and $i^{+}$, the future timelike infinity, as end of its trajectory. Both timelike infinities are points (or vertexes) in the Carter-Penrose diagram. 

It is worth to emphasize that the difference among the motions of bodies is due to the mass of bodies in general relativity. Photons, the particles of light, are massless particles. Then, their \enquote{natural places} are $\mathscr{I}^{\pm}$. For observes (massive particles), who have positive mass, $i^{\pm}$ are their \enquote{natural places}. The hypothetical tachyons would have imaginary mass, and then would travel along spacelike geodesics and violate causal relations. Thus, their \enquote{natural place} would be $i^{0}$. In Minkowski spacetime, it is straightforward to relate the mass of particles and velocities. Then, massless particles travel at the speed of light, massive particles travel slower than the speed of light, and particles with \enquote{imaginary} mass would travel faster than the speed of light.

Infinities as natural places suggest an analogy. Like every analogy, it has problems, or limitations. Considering the six Aristotelian requirements to the notion of place, we may investigate whether or not the notion of conformal infinities is suitable to the notion of natural place. At first glance, the five infinities ($\mathscr{I}^{\pm}$, $i^{\pm}$, and $i^{0}$) from the Carter-Penrose diagrams may obey the requirements (2), (4) and half of (6). Other requirements would not be satisfied because we are talking about infinities, then a body supposedly would not reach the natural place. Moreover, the requirement \textit{rest at the natural place} would present problems as well. If a notion of place is totally possible in general relativity (in spacetimes occupied by particles), a notion of natural place will be problematic because the Aristotelian notion postulates rest at the natural place. In particular, for null geodesics, photons tend to the future null infinity $\mathscr{I}^{+}$ with speed of light. According to the second postulate of special relativity, rest is forbidden for massless particles. But in an infinite spacetime any position of an observer may be considered infinitely distant from other observers. Moreover, for timelike geodesics in Minkowski, rest is possible, for example, for a body or observer at $r=0$. Such a body remains at rest. Then, for bodies with positive mass (maybe the unique case described by Aristotelian physics and their elements), the notion of natural place works. Observers at rest are at their \enquote{natural places}. However, the notions of \enquote{above} and \enquote{below} are nonsense in an infinite spacetime such as the Minkowski solution. The requirement of absolute notions of \enquote{above} and \enquote{below} is not fulfilled in infinite spaces. But this was already indicated by \citet[$295b5$]{Heavens}.      

Above all, the most important analogy is found in the motion that tends to infinity through geodesic motion. The movement to the conformal infinities depends on the body property, i.e., its mass. In Aristotelian physics, the same dependence on the body properties is noted during the natural motion.

\subsection{Friedmann-Lemaître-Robertson-Walker spacetime}

Minkowski spacetime is a vacuum solution of Einstein's field equations, i.e., it is an empty world, or, approximately, it is a flat and static spacetime with tiny particles. Then, an expanding cosmological geometry with matter content is more appropriate to think about natural places in a \enquote{real} world. In Einsteinian cosmology, the standard metric is the Friedmann-Lemaître-Robertson-Walker metric, which in the $t$, $r$, $\theta$, and $\phi$ coordinates reads 
\begin{equation}
ds^2 =g_{\mu\nu (FLRW)}dx^\mu dx^\nu =  - dt^2 +a(t)^2\left[ \frac{dr^2}{1-kr^2}+ r^2 \left( d\theta^2+\sin^2\theta d\phi^2  \right) \right],
\label{FLWR} 
\end{equation}
where $a(t)$ is the scale factor, which indicates, for example, the rate of expansion of the spacetime fabric, and $k$ is the spatial curvature in the universe. The recent data \citep{Planck} point out to an expanding universe and $\vert \Omega_{k}\vert <0.005$ for the density parameter of the spatial curvature, i.e., for this value of $\Omega_{k}$ the universe is almost flat ($k\approx0$).  The coordinate ranges in the cosmological metric are $0< t<\infty$, $0\leq r <\infty$, $0 \leq \theta \leq \pi$, and $0 \leq \phi \leq 2\pi$. Following \cite{Hawking}, it is more appropriate rewrite Eq. (\ref{FLWR}) using the conformal time $\tau$, which is defined as $d\tau = dt/a(t)$, and a new radial coordinate, $\chi = \arcsin r$. Then, the standard metric becomes
\begin{equation}
ds^2= g_{\mu\nu (FLRW)}dx^\mu dx^\nu = a(\tau)^2 \left[- d\tau^2+d\chi^2+f(\chi)^2\left(d\theta^2+\sin^2\theta d\phi^2 \right)  \right],
\end{equation}
with $f(\chi)=\chi$ for a spatially flat universe. 

\begin{table}[t]
  \centering
  \begin{tabular}{ | c | l | c | }
    \hline
    \textbf{\footnotesize Carter-Penrose Diagram} & \textbf{\footnotesize Origin of the Trajectories} & \textbf{\footnotesize End of the Trajectories}\\ \hline
    
    \begin{minipage}{6cm}
     \includegraphics[width=\linewidth, height=40mm,trim=50 450 0 0]{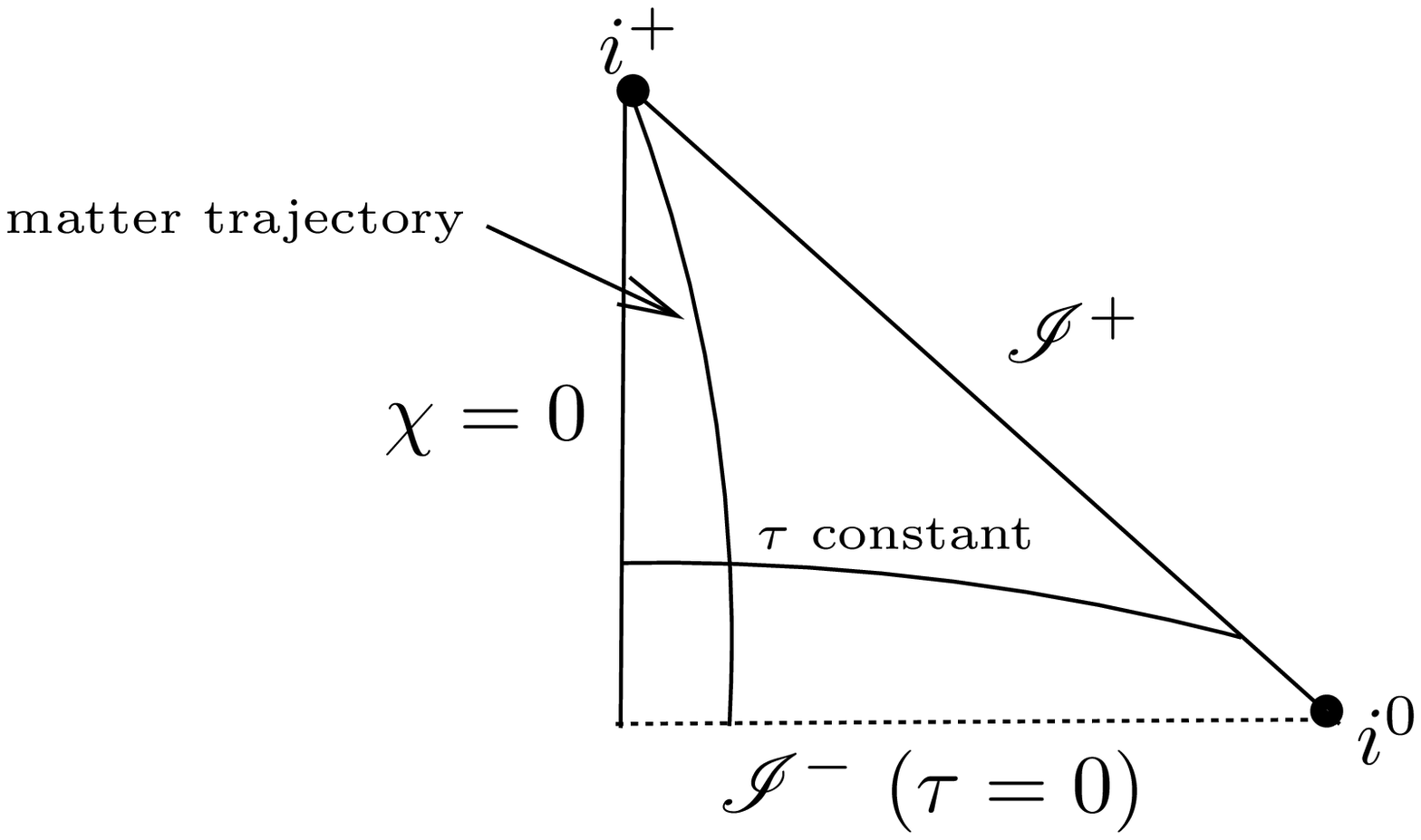}
    \end{minipage}
    &
      \begin{minipage}{3.45cm}
      \begin{itemize}
       \item {\scriptsize $\mathscr{I}^{-}$ (the big bang) }
       \end{itemize}
    \end{minipage}
    & 
     \begin{minipage}{4.5cm}
      \begin{itemize}
        \item {\scriptsize $i^{+}$  (for timelike geodesics)}
        \item {\scriptsize $\mathscr{I}^{+}$ (for null geodesics)}
        \item {\scriptsize $i^{0}$ (for spacelike geodesics)}
      \end{itemize}
    \end{minipage}
    \\ \hline
  \end{tabular}
  \caption{Carter-Penrose diagram for expanding Friedmann-Lemaître-Robertson-Walker spacetime with $k=0$ (spatially flat), matter content described by a pressureless perfect fluid, and without a cosmological constant. The conformal infinities ($i^{+}$, $\mathscr{I}^{+}$, and $i^{0}$) indicate the end for the trajectories of bodies. The future timelike infinity $i^{+}$ and spatial infinity $i^{0}$ are points in the diagram. On the other hand, the infinities $\mathscr{I}^{\pm}$ are lines, and in particular the infinity $\mathscr{I}^{-}$, described by the traced line, is the origin of everything because it indicates the big bang, or the initial singularity at $\tau=0$. The point $\chi=0$ represents the origin of the coordinate system $\tau$, $\chi$, $\theta$, and $\phi$. As we can see, the cosmological diagram is half of the Minkowski diagram, and the notion of future infinities is identical.}
\label{Diagram_FLRW}
\end{table}

The same procedure for generating the Carter-Penrose diagram is possible for the cosmological metric. As we can see, for $k=0$, the standard metric is conformal to Minkowski space (assuming that $\chi$ is the radial coordinate $r$).  Then, such as the Minkowski's geometry, the cosmological standard metric is conformal to the Einstein static universe by using the coordinate transformations (\ref{Transformations}) (assuming $\tau=t$ and $\chi=r$). That is, 
\begin{equation}
g_{\mu\nu (E)}=\Omega^{'2} g_{\mu\nu (FLRW)},
\label{E_FLRW}
\end{equation}  
with $\Omega^{'2}=(\cos T +\cos R)^2/a(\tau)^2$ playing the role of the conformal factor in this case. However, for the flat universe ($k=0$) with matter\footnote{In cosmology, matter means a pressureless perfect fluid.} and without a cosmological constant, the Einstein static metric used in the relation (\ref{E_FLRW}) has limited time coordinate: $0<T<\infty$. Such as in Minkowski spacetime, infinities in the standard metric are boundaries in an open region of the Einstein static universe. But the Carter-Penrose diagram for the standard metric in cosmology does not present the same structure of infinities of Minkowski spacetime. The difference is $\mathscr{I}^{-}$, which in the standard metric describes the origin of all particles and is not the past null infinity of Minkowski spacetime. In Friedmann-Lemaître-Robertson-Walker spacetime, $\mathscr{I}^{-}$ is a spacelike surface, the supposed big bang,\footnote{Today it is possible to construct cosmological models without the initial singularity in the general relativity context. Such regular models are the so-called bouncing cosmologies \citep{Novello,Neves2}. Contrary to Minkowski spacetime, the cosmological metric in the standard model is not geodesically complete. In the past, the initial singularity is the end of trajectories. In Minkowski spacetime, as we have already seen, we have a structure of past infinities. This is the reason why Minkowski spacetime is geodesically complete and the Friedmann-Lemaître-Robertson-Walker metric is not.} or the initial singularity at $\tau=0$. According to the standard model, it is the origin of the universe and the beginning of the matter trajectory. On the other hand, the end of the trajectories, the structure of future infinities ($\mathscr{I}^{+}$, $i^+$, and $i^0$), is identical to the  Minkowski's solution (see Table \ref{Diagram_FLRW}). 

In the standard cosmological metric, as well as in the Minkowski metric, the notion of \enquote{natural places} is suggested. However, if we consider an expanding universe, rest with respect to the surrounding bodies will be impossible even for bodies along the timelike trajectories. Thus, the Aristotle's exigence of rest at the natural place is not fulfilled for any trajectory in the cosmological scenario. But the movement to infinities, or \enquote{natural places,} such as in Minkowski spacetime, is guided only by a body property, its mass, which is due to its \textit{arche}.

\section{Final remarks}
In many aspects, ancient science---specifically its models but not all ideas---is ruled out today. The geocentric model and the teleological point of view, for example, should be found only in history books. However, ancient thought may offer tools and perspectives to look at problems today. In ethics, Greek philosophy is full of contributions to think about or to discuss values. In the arts, the Aristotelian art criticism is useful for writers, poets, and enthusiasts. In science, specifically in physics, it is possible to find similarities and analogies. In many cases, like the concept of energy (\textit{energeia} in Greek) and potential energy, physicists are unaware of the influence of ancient thought on modern science. In this article, I showed that the concept of natural place within Aristotelian physics has an analog in general relativity. The conformal infinities in spacetime diagrams, also called Carter-Penrose diagrams, provide a notion of \enquote{natural place} within Einsteinian theory. The three types of geodesics---timelike, null and spacelike---lead to the concepts of timelike infinity, null infinity, and spacelike infinity as tendency, or \enquote{places,} for bodies along geodesics. The analogy is not perfect like every analogy. The Aristotle's requirements for the notion of natural place are not completely fulfilled for conformal infinities. But some aspects are retained by the Carter-Penrose diagrams: the movement to a \enquote{place} for bodies along geodesics and the dependence on the body's property (its mass) to specify the geodesic (or \enquote{natural}) motion. 

\section*{Acknowledgments}
I would like to thank IMECC-UNICAMP for the kind hospitality.

\end{document}